\documentstyle[aps,multicol,epsfig,psfig,pre,colordvi]{revtex} 
\tightenlines 
\begin{document} 
  
\title{Dynamical scaling of the DNA unzipping transition } 
\author{Davide Marenduzzo$^{1}$,Somendra M. Bhattacharjee$^{2,3}$, 
  Amos Maritan$^{1,4}$, Enzo Orlandini$^{3}$ and Flavio Seno$^{3}$ } 
\address{$^1$ International School for Advanced Studies (SISSA),\\ 
  and Istituto Nazionale di Fisica della Materia,\\ 
  Via Beirut 2-4, 34014 Trieste, Italy \\ 
  $^2$ Institute of Physics, Bhubaneswar 751 005, India \\ 
  $^3$ INFM- Dipartimento  di Fisica - Universit\`a di Padova - Italy\\ 
  $^4$ The Abdus Salam International Center for Theoretical Physics 
  (ICTP),\\  
  Strada Costiera 11, 34100 Trieste, Italy  } 
 
\maketitle 
\begin{abstract}  
We report studies of the equilibrium and
the dynamics of a general set of lattice models
 which capture the essence of
the force-induced or mechanical DNA unzipping transition.
 Besides yielding the 
whole equilibrium phase diagram in the force vs temperature plane,
which reveals the presence of an interesting re-entrant 
unzipping transition for low $T$, these models enable us 
to characterize the dynamics of the process starting from a 
non-equilibrium initial condition.
 The thermal 
  melting of the DNA strands displays a model dependent time evolution.
On the contrary, our results suggest
 that the dynamical mechanism for the unzipping by force
is very robust and the scaling behaviour does not
 depend on the details of the description we adopt.
\end{abstract}
  
\begin{multicols}{2} 
   
The replication of DNA is a correlated process involving many 
proteins and other molecules\cite{kornberg} working at different 
points in space and time. An understanding of the nature and origin 
of this correlation is expected to shed light on this complex 
mechanism.  It has recently been shown\cite{smb,sebastian,lubnel,zhou,antonio}
 that the force 
induced unzipping of DNA is a genuine phase transition different 
from the thermal melting transition of DNA.  It was then 
hypothesized \cite{smb} that the initiation of replication at the origins along 
the DNA, e.g, by dnaA for E.Coli\cite{kornberg,dnaa} or by the 
``origin recognition complex'' (ORC) in eukaryotes\cite{orc} is like 
 this unzipping near the critical threshold (with dnaA or ORC acting 
as the force-inducing agent) and the resulting correlation during 
unzipping leads the co-operativity required for replication. 
   
In contrast to real biological situations, techniques like laser 
tweezers\cite{heslot}, atomic force microscopes 
(AFM)\cite{reif,anselmetti,strunz} etc have been used to study DNA 
by pulling at one end.  This has led to strand separation by force. 
In particular, AFM experiments reported hysteresis in the unzipping 
process, indicating the presence of a first order 
    transition.  These mechanical unzipping experiments have opened 
up new ways of thinking about DNA, just as similar stretching 
experiments of DNA showed the possibility of several structures other 
than the most prevalent B-DNA\cite{allemand}.  The activities of 
polymerases, topoisomerase etc on single stranded DNA have now been 
analyzed in terms of the force they exert or the force applied against 
them\cite{force,force2,force3}.  What needs to be investigated, to 
mimic the biological situation, is the coupling between the opening 
of the strands and the subsequent events during replication.  Such a 
study involves the dynamics of the unzipping process\cite{sebastian}. 
  
The purpose of this paper is to define a set of simpler models, in the 
spirit of Poland and Sheraga\cite{PS66}, for which the  
unzipping transition can be studied exactly.  On the basis of this, the 
dynamics can be understood. The proposed lattice models  (bubble 
models: b-models) 
 incorporate the mutual-avoidance (hard-core repulsion) 
 of the strands (and also 
self-avoidance).  A further simplification is obtained by suppressing 
the bubbles along the chains, thereby defining a ``fork model'' 
or ``Y-model''. The 
phase diagram of the equilibrium system displays in both cases a  
re-entrant region at low $T$: for a finite 
range of forces the molecule gets unzipped by {\em decreasing} 
the temperature. The 
dynamics  of both the b- and the Y-models 
 in the various phases and on the phase boundary are then 
studied, by starting from a zig-zag non-equilibrium bound state as the 
initial condition. We find that in all regimes above or below the critical
line in the phase diagram, the time evolutions of the order parameters 
follow dynamical scaling laws.
  
\noindent {\bf The models and their equilibrium phase diagram:} 
We model the two strands of DNA by two directed self and mutually 
avoiding walks.  In two dimensions, on the square lattice (see Fig.1), 
 this means 
that the two walks are forced to follow the positive direction of the 
diagonal axis $(1,1)$ (i.e. the coordinate along the direction 
$(1,1)$, to be called the $z$ direction, always increases). 
The force is then acting along the direction perpendicular 
to direction $z$, i.e.,\ \ $(-1,1)$: this transverse 
direction is called the $x$-direction.  By measuring the $x$ 
separation in unit of the elementary square diagonal, we say that two 
complementary monomers are in contact when $x=1$: a (positive) 
    binding energy $\epsilon$ is gained for each contact.  Due to 
the geometrical properties of the lattice, all these contacts 
contributing to the energy involve monomers labelled by the same 
$z$-coordinate, as one would require for base pairing in DNA (see Fig. 
1). 
 
\vbox{ 
\begin{figure}   
\centerline{\psfig{figure=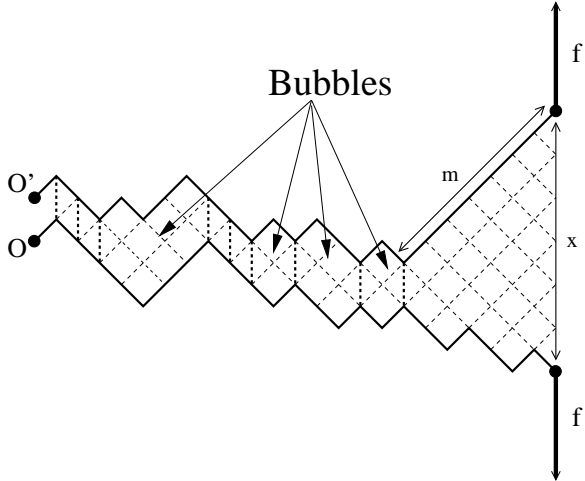,width=3.in}}  
\narrowtext 
\caption{ A typical configuration of the two DNA-strands  
as modelled on a square lattice. Dashed lines indicate 
monomers which are in contact. The quantities $m$ and $x$ are 
graphically represented.  
} 
\label{fig:1} 
\end{figure} 
} 
 
By fixing $\epsilon= +1$ 
throughout the calculations, all the thermodynamic properties of the 
system depend on the temperature $T = \beta^{-1}$ and on the force 
$f$. We have set $k_B=1$. The partition function of the two $n$-step 
chains, one starting at $(0,1)$ and the other at $(1,0)$ is 
\begin{equation}                 
{\cal Z}_n(\beta,f) = \sum_{x \geq 1} d_n(x) \exp{\left( \beta 
 fx\right)} 
\sim e^{-n\beta{\cal F}}, 
\label{eq:eq5} 
\end{equation} 
where $d_n(x)$ represents the fixed distance partition function, {\em i.e.}, 
the sum over all interacting pairs of directed chains whose last 
monomers are at distance $x$\cite{nota1} and ${\cal F}$ is the free 
energy density.  The quantity $d_n(x)$ whose $\beta$ dependence has 
been omitted, obeys simple recursion relations.  In $d=2$, on the 
square lattice, the recursion relation is 
\begin{eqnarray} 
d_{n+1}(x) &=& \left[ 2 d_n(x)+d_n(x+1)+d_n(x-1) \right ]  \nonumber  \\ 
&&\quad\quad\times \left[ 1 + \left( e^{\beta} - 1 \right) 
  \delta_{x,1} \right ],  
\label{eq:eq6} 
\end{eqnarray} 
valid for $x \geq 1$ with the conditions $d_0(x)=\delta_{x,1}$ 
and $d_n(0)=0$ for all $n$.  Note that this kind of equation appears 
also in other models of DNA\cite{PS66,Causo} and in studies of random 
walks adsorption\cite{Rubin1}.  The model can be asymptotically solved 
by locating the singularity closest to the origin of its related 
generating function\cite{Lifson} 
\begin{equation} 
{\cal G}(\beta,f) = \sum_{n=0}^{\infty} z^n {\cal Z}_n(\beta,f). 
\label{eq:eq7} 
\end{equation}        
A phase transition is indicated by any singular change in the location of the
 closest singularity.

With some simple calculations a critical line, 
$f_c(T)$, separating 
the zipped from the unzipped phase can be obtained 
\begin{equation} 
f_c(T)=T \cosh^{-1}{ 
\left[ 
\frac{1}{2} 
\frac{1}{\sqrt{1- e^{-\beta}}-1 + e^{-\beta}}  -1 
\right], 
} 
\end{equation} 
\noindent and  it  has been plotted in Fig. 2. 
  
\vbox{ 
\begin{figure}   
\centerline{\psfig{figure=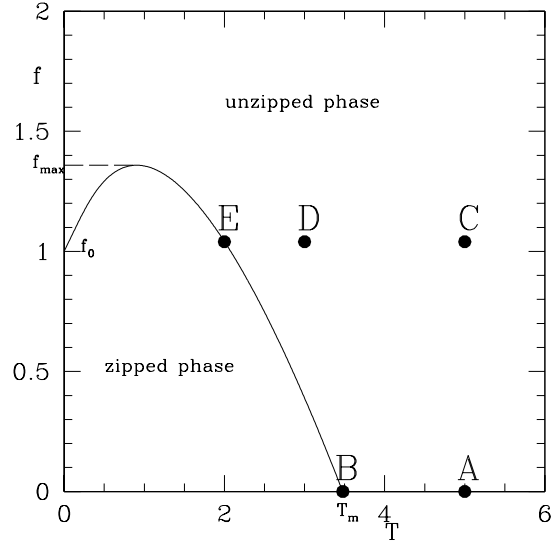,width=3.in}} 
\narrowtext 
\caption{The plot of the force vs. 
  temperature phase diagram for the model on the directed lattice.} 
\label{fig:histo} 
\end{figure} 
} 
 
At $f=0$ the critical melting temperature is 
$T_m=\frac{1}{\log{4/3}}$, the critical force at $T=0$ is $f_0=1$ 
whereas for $f > f_{\rm max}=1.358806...$ the system is always in the 
unzipped state. Similar results can be obtained for the  
physically relevant three dimensional case and for higher  
 dimensions\cite{antonio}. 
The average fraction of contacts $\Theta$ 
is zero (not zero), when $f > f_c(T)$ ($f <f_c(T)$) while the average  
end-to-end distance $\langle x\rangle= - N \frac{\partial 
  {\cal F}}{\partial  f}$, where $N$ is the number of base pairs
in each chain, 
  and the average number of 
 ``liberated'' monomers ({\em i.e.} from the last contact to the 
end), $m$ (See Fig. \ref{fig:1})  have the opposite behavior. The 
transition line is first order at any point with $f_c(T) \neq 0$ in 
$d=2,3,4,5$ and everywhere for $d > 5$.  
  
We notice that there is a re-entrance in the $f-T$ phase diagram. For 
$f_{\rm max} > f > f_0$ the usual denaturation transition is present 
but if the temperature is further lowered the two strands separate 
again through a ``cold unzipping''\cite{saws,unpub}. 
   
A further simplification to this model can be obtained by suppressing 
the bubbles along the chain, {\em i.e.} by considering only 
conformations having the first $N-m$ monomers bounded, whereas the 
remaining $m$ are separated in a Y-like conformation. 
  
This Y-model will be extremely useful to study the unzipping 
dynamics and it presents a phase diagram similar to the one previously 
obtained but, for example, with $T_m=\frac{1}{\log{2}}$, $f_0=1$ and $ 
f_{\rm max}=1.282143...$\cite{note2} in d=2. 
 
\noindent{\bf Dynamics.} 
We now consider the dynamics of the  models  
 previously introduced, the b-model of directed 
 walks and the simplified 
Y-model. In both cases, we start from a non-equilibrium 
``zero-temperature'' initial condition with the two chains zipped in a 
zig-zag configuration, and let the system evolve at a temperature $T$ 
and under a force $f$, with $T$ and $f$ chosen so that the equilibrium 
state is either on or above the critical line. 
The simpler 2-dimensional case will be presented first and the 
generalization 
to higher dimensions will be discussed later. 
 
  The five regimes 
considered are marked A-E in Fig. \ref{fig:histo}. 
Numerically,  a Monte-Carlo dynamics is used to   
monitor the time evolution of the order parameters $m$ and $x$ 
previously introduced (see also Fig. 1).  
For the dynamics,  we selected only local physical moves, so that the 
model should be the  
lattice counterpart of the Rouse model in the continuum. In all cases we find 
that, far from saturation, the order parameters evolve as power laws as functions of time $t$: 
in particular we can define two exponents $\theta_1$ and $\theta_2$ 
as  
\begin{equation} 
  \label{eq:1} 
m(t)\sim t^{\theta_1},\quad {\rm and}\quad  x(t)-x(0) \sim 
t^{\theta_2},   
\end{equation} 
 Notice that $m$ and $x$ in Eq. (\ref{eq:1}) refer to the ensemble 
 average values, but  average signs  have been omitted for 
simplicity of notation. We in fact find the following  dynamical 
scaling laws to hold: 
\begin{equation} 
 \frac{m(t)}{N^{d_1}}  =   G_m\left(\frac{t}{N^{z_1}}\right) {\rm , \,} 
 \frac{x(t)-x(0)}{N^{d_2}} =   G_x\left(\frac{t}  
{N^{z_2}}\right),\ 
\label{scaling} 
\end{equation} 
where $N$ is the length of the chains and $ G_{m,x}$ are two scaling 
functions. Eq. \ref{scaling}  also defines the exponents $d_{1,2}$ and 
$z_{1,2}$ for the two variables.  Note that $\theta_1=\frac{d_1}{z_1}$, 
$\theta_2=\frac{d_2}{z_2}$ and that $d_{1,2}$ can be obtained through 
equilibrium considerations as one requires 
$m(t)\stackrel{t\to\infty}{\sim}N^{d_1}$ and 
$ x(t) \stackrel{t\to\infty}{\sim}N^{d_2}$.   
The crossover to the equilibrium behavior is described  by the 
``dynamic'' exponents $z_{1,2}$.  The possibility of the two 
quantities having  different relaxations is kept open by  two 
different exponents.

The values of the exponents in Eqs. (\ref{eq:1}) and (\ref{scaling}) 
obtained from simulations, for both the Y-model and the b-model, are 
shown in Table I.  They were obtained by collapsing Monte-Carlo data 
according to eq. (\ref{scaling}) and by using a recently proposed 
search algorithm \cite{BS01} (see Fig. 3) 
 
\begin{figure} 
\centerline{\psfig{figure=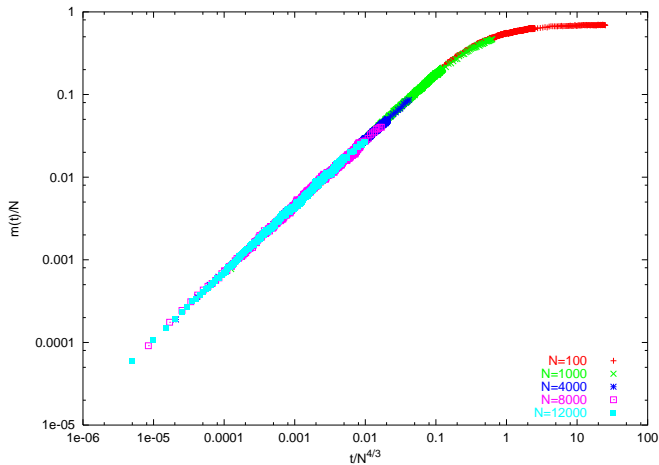,width=3.5in}} 
\narrowtext 
\caption{Plot of $\frac{m(t)}{N}$ vs  
$\frac{t}{N^{\frac{4}{3}}}$ for 
various values of $N$: the collapse of all curves indicates that 
$d_1=1$, $z_1=\frac{4}{3}$ and $\theta_1=\frac{3}{4}$. } 
\label{collapse} 
\end{figure}

It is possible to explain  the exponents for the 
Y-model found numerically. There are two  relevant mechanisms 
which can drive the separation of the strands: 
the unbinding of bases at the bifurcation point and the 
 stretching of the strands at the extremes. The combination of these two 
processes controls the overall  behavior in different regimes.

{\em Regime A: $f=0$, $T>T_m$}: Above 
the critical temperature 
the dominance of the  entropy implies that at every time step one base 
pair breaks, yielding a linear 
behavior with $\theta_1=1$ and $d_1=1$. Also $x(t)$ tends 
 to increase, up to its equilibrium value 
$N^{\frac{1}{2}}$: this is reflected in the equilibrium probability 
distribution $P_{eq}(x)$ which displays an upward derivative at $x=1$. 
This suggests (see \cite{barabasi})  
that the dynamics of this quantity should be in the same 
    universality class of 
the one-dimensional Kardar-Parisi-Zhang equation\cite{barabasi}, and so $\theta_2=\frac{1}{3}$, 
$d_2=\frac{1}{2}$, and $z_2=\frac{3}{2}$.

{\em Regime B: $f=0$, $T=T_m$}: In this regime 
$\theta_1=\frac{1}{2}=z_1^{-1}$ because at criticality the 
probabilities to increase and to decrease $m$ are expected to be 
equal, so that $m$ performs, roughly speaking, a random walk in time 
with reflecting boundaries at $m=0,N$. Also for the end-to-end 
distance, steps toward larger or smaller values of $x$ are equally 
probable, and this means that the equation obeyed by $x$ should stay 
in the same universality class as the one-dimensional
Edward-Wilkinson equation, and 
therefore $\theta_2=\frac{1}{4}$ and $z_2=2$. 
 
{\em Regime C: $f>0$,$T>T_m$}: The strands microscopically tend to 
stretch along the direction of the pulling force. However, once we 
have pulled the two chains up to an end-to-end distance $x$, to 
increase $x$ further by one unit we first need to move all the stretched part, 
which would take a time typically of order $x$. In other words, one has 
$x(t+t_0)\sim x(t)+\frac{t_0}{x}$ and we expect the dynamical exponent 
$\theta_2$ to be  $\frac{1}{2}$ (of course $\theta_1=1$ as before and 
$d_1=d_2=1$).  
 
{\em Regime D: $T<T_m$,$f>f_c(T)$}: Here the only microscopic 
mechanism for opening the fork is through the applied force: the 
strands must stretch completely in the vicinity of the bifurcation 
point and only at this point will the fork liberate one more monomer, 
because otherwise the opening is energetically very unlikely.  Thus we 
expect that $x\sim m$ and, using arguments as done for regime $C$, 
$\theta_1=\theta_2=\frac{1}{2}$, $d_1=d_2=1$.  In ref.\cite{sebastian} 
it was found, in the mean field approach and in a model resembling our 
Y-model, that the time necessary to unzip the two strands 
completely is of the order of $N^2$. This is in agreement with our 
analysis, but works only in this regime. 
 
{\em Regime E: $T<T_m$,$f=f_c(T)$}: At criticality, one expects that 
the cost for unzipping and zipping is the same (the equilibrium 
probability distribution of having $m$ monomers unzipped or an 
end-to-end distance equal to $x$ is flat), therefore $x(t+t_0)=x(t)\pm 
\frac{t_0}{x}$ with equal probability. Therefore, the end-to-end 
distance makes a random walk in the rescaled time $\frac{t}{x}$ so 
that $x\sim(\frac{t}{x})^{\frac{1}{2}}$ and 
$\theta_2(=\theta_1)=\frac{1}{3}$ and $d_1=d_2=1$ since at coexistence 
there is a finite fraction of liberated monomers. 
 
Another way of obtaining $\theta_1=\frac{1}{3}$ is to demand that a kink 
liberated at the fork needs to diffuse out of the end before the next 
one is released. In other words, the rate of change of $m$ is 
determined by the diffusion of a kink over a distance $m$.  The latter 
time-scale being of order $m^{-2}$, we expect $dm/dt \sim m^{-2}$ which 
gives $m \sim t^{\frac{1}{3}}$. 
 
This simple model deserves some observations.  Remarkably, it displays 
two different time scales in its dynamics, mirrored in the difference 
of the exponents $z_{1,2}$ in regimes A and C.  The first time scale 
($t\sim N^{z_1}$) quantifies the time necessary for the unbinding (or 
unzipping) of the bases while the second ($t\sim N^{z_2}$) gives the 
time needed to open (and to stretch whenever $f\ne 0$) the two chains 
up to their equilibrium end-to-end distance. At $T < T_m$ the two 
processes are virtually the same, because the unbinding (or unzipping) 
is dragged by the stretching.  However, above $T_m$, the processes 
decouple and the unbinding gets faster, being controlled by the 
temperature, and therefore $z_1<z_2$. 
 
Furthermore, in the numerical calculations we found large 
sample-to-sample fluctuations, thereby requiring a huge number of runs 
to get good averages of 
$m(t)$ and $x(t)$. This is due to the long time correlation that 
exists in the system, which keeps samples with different initial 
histories far apart for any $t$.

Turning to the b-model with bubbles, amazingly the dynamical 
exponents in regimes B, E and D with $T\le T_m$ are the same as found 
in the Y-model. This establishes that at $T<T_m$ not only for 
statics, as we saw in the previous section, but also for the dynamics, 
bubbles are not relevant in the scaling properties (for $T=T_m$ the 
equality is due to another reason \cite{not3}).  At $T\ge T_m$, on the 
other hand, the opening of bubbles heavily affects the base unpairing 
process, unlike the Y-model case where bubbles are forbidden.  The 
length of the unzipped part in the present case now can change by $\pm 
l(t)$, where $l(t)$ is the typical length of bubbles, and the motion 
of the fork point can by no means be approximated by a simple random 
walk ( and so $\theta_1$ changes as shown in Table I). The quantity 
$x(t)$ instead has a dynamics in the b-model similar to the fork 
case and indeed $\theta_2$ is the same for both the models in all 
regimes.  We show in Fig.3 the collapse leading to 
$\theta_1=\frac{3}{4}$ in regime $A$.

An important question is that of the dependence of these results on 
dimensionality.  As for the Y-model, the arguments we gave suggest 
that there be no $d$-dependence. For the b-model, instead, this 
should be true only in regimes D and E for $T<T_m$, where the  
Y-model gives the exact result; at $T\ge T_m$, on the other hand, bubbles 
play a dominant role and so we expect a dependence on $d$.  We 
confirmed this picture with some calculations on a simpler model which 
should be in the same universality class of the one under study: 
that of a single random walk, pinned at the origin by an attractive 
interaction and subject to a stretching external force. In this system 
$m$ is defined as the number of monomers from the last visit to the 
origin to the end of the walk. For $T>T_m$, in regimes $A$ and $C$, 
our calculations show that the exponent $\theta_1$ increases as 
dimension increases, apparently with no upper critical dimension. Just 
at criticality at zero force, instead, we find that the exponent 
$\theta_1$ is very close to $\frac{1}{2}$ in any $d$.  The emerging picture 
of robust results for $T<T_m$ and model-dependent dynamics for $T\ge T_m$ would 
be preserved even if,  in the original models, 
 the directedness constraint is relaxed (as is the case 
for the statics\cite{unpub}). 
 
The arguments presented so far could be generalized to add other 
ingredients of dynamics also.  An example is the nonlocal effects 
 in dynamics, as is the case {\em e.g.} 
in the Zimm model.  In ref.\cite{sebastian}, the author suggests that, 
in the regime we call $D$, in the mean field approximation, 
non-locality could speed up the opening so that $\theta_{1,2}$ would 
be $\frac{1}{1+u}$ instead of $\frac{1}{2}$ ($u$ is the exponent 
characterizing the length dependence of the mobility as defined, e.g., 
for the Zimm model, see \cite{sebastian} for the notation and 
\cite{edwards} for a review).  In regime $E$, we can combine our 
arguments with the same reasoning to get 
$\theta_1=\theta_2=\frac{1}{2+u}$.  As we expect, the Rouse model 
results are obtained with $u=1$. 
 
Lastly, the above analysis can also be extended to binary disorder in 
the contact potential, which is of course a realistic feature to be 
included in the model, {\em i.e.}  the energy of the contact of the 
$i$-th base may be $\epsilon_i=\epsilon+\Delta_i$, $\Delta_i$ being a 
random variable with binary distribution and zero mean.  The  
Y-model offers a good starting point for the study of the effects of 
heterogeneity, because the critical line of the quenched model  
can be proved to be  the same as that of the pure model with energy 
$\epsilon$.  
If we call ${\cal F}(m)$ the free energy density of a configuration 
with $m$ (out of $N$) unzipped monomers, we expect on general grounds 
that $P(m,t)$, the probability of having $m$ monomers unzipped at time 
$t$, will obey a master equation with transition rates $W_{\pm,m}$ 
depending on the realization of disorder, {\em i.e.} we expect that 
$W_{\pm,m}\propto{\rm min} \left\{e^{\mp \beta N\frac{d{\cal  F}(m)}{dm}},1\right\}$,  
with $N\beta\frac{d{\cal F}(m)}{dm}$  
$={+\frac{\Delta_{N-m}}{T}+\frac{\epsilon}{T}-  \log(1+\cosh(\frac{f}{T}))}$ 
\cite{note4}, which can be seen to 
consist of a zero-mean random ``noise'' ($\frac{\Delta_{N-m}}{T}$) 
plus a ``bias''.  We thus believe that the dynamics of the Y-model 
can be mapped onto Temkin's model of a random walker in a random 
environment \cite{hugh} (provided that at $T<T_m$ we rescale     
time  as described above  in regimes $D$ and $E$ ), 
so that, following Ref. \cite{derrida}, 
there will be a region around the critical line, at $T<T_m$, 
in which $m,x\ll t^{\frac{1}{2}}$ (and just on the critical line at $f\ne 0$ 
one is tempted to expect 
$m,x\sim\left(\log\left(\frac{t}{x}\right)\right)^2$).  
The quantities $m,x$ are now quenched 
averages over realizations of disorder. The scenario we propose is 
sketched in Fig.4.  The curve bounding the region 
where, at $T<T_m$, 
 disorder should be important for the dynamics has been found by applying 
the criterion of Ref. \cite{derrida} to the system with the above 
transition rates. This curve coincides 
 with the critical line of the static model 
with annealed disordered. 
Preliminary runs at very low $T$ suggest that  
as $T\to 0$, for a given realization of disorder and 
 with any force between $\epsilon_1$ and 
$\epsilon_2$ (the two binding energies), the unzipping will take 
place as far as the first more attractive base pair is found. 
Besides, the probability that the monomer of index $m_0$ will ever be 
unzipped decays exponentially: $P(m_0)\propto 2^{-m_0}$. 
 This is consistent with the mapping 
suggested above. A detailed study of the effect of disorder  
will be done elsewhere.

\noindent{\bf Conclusion.} 
We obtained the phase diagram of a lattice model for the unzipping 
transition of a double stranded DNA by a force.  This model 
incorporates both the mutual and self-avoiding nature of the two 
strands and the equilibrium problem can be solved exactly.  A still 
simpler model has been defined by suppressing the bubbles that are 
important for thermal denaturation.  This simpler version is also 
exactly solvable and is shown to retain the basic features of the  
b-model.  Thanks to the exact knowledge of the phase boundary, we have 
been able to investigate the dynamics of unzipping from a 
non-equilibrium bound state both on and away from the phase boundary. 
The dynamics shows scaling behaviors in different regimes of the 
phase diagram.  These scalings in most cases could be understood from 
the plausible mechanisms of unzipping and denaturation as  
discussed in the previous sections.  We end with two notable features. 
In the case of the b-model with bubbles, the unzipping at high 
temperatures remains a puzzling issue, especially the dynamic exponent 
$z_1=\frac{4}{3}$ with or without force. We believe that this has to 
do with the statistics of bubbles, though no satisfactory answer could 
be found.  This remains an open problem.  Lastly, on a positive note, 
the unzipping dynamics on the phase boundary in the presence of a 
force is found to be distinctly different from the thermal 
denaturation at zero force.  Whether a real biological system takes 
advantage of these differences to distinguish the unzipped region of 
DNA from a fluctuation-induced bubble formation remains to be probed. 
\begin{figure} 
\centerline{\psfig{figure=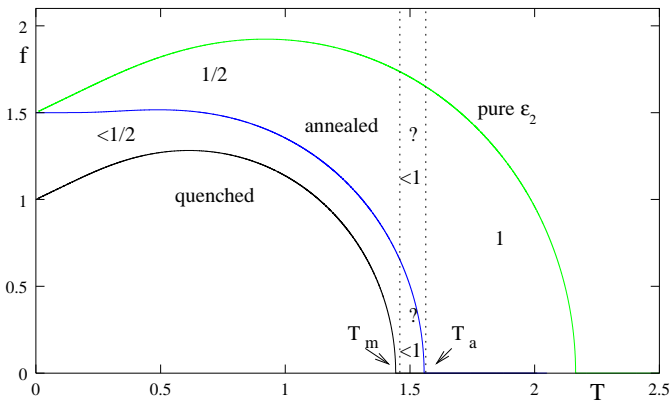,width=3.5in}} 
\narrowtext 
\caption{Plots of the phase diagram for homo-DNA (top) made up of the more 
attractive base ($\epsilon_2>\epsilon_1$), and for the annealed 
(middle) and quenched (bottom) disordered models. The numbers refer to 
the values of the exponent $\theta_1$ in the various regions of 
the phase diagram, 
which we believe to hold on the basis of the mapping onto 
Temkin's model. Note that above the annealed critical line  at $T<T_m$  
and everywhere at $T>T_a$ 
the pure system results are recovered. Interestingly, 
the annealed line does not  
show the re-entrant behavior found in Section II and in the  
quenched system. ``?'' indicates the regime 
 where we do not have numerical evidence in support of our results. 
We took $\epsilon_2=1.5$ and 
$\epsilon_1=0.5$ in the calculations.} 
\label{scenario} 
\end{figure} 
\vbox{    
\begin{table}[htbp] 
\begin{center} 
\begin{tabular}{| c || c | c | c || c | c | c |} 
 
 Regime & $d_1$ & $z_1$ & $\theta_1$ & $d_2$ & $z_2$ & $\theta_2$\\ 
\hline \hline 
A:Y & $1$ & $1$ & $1$ & $1/2$ & $3/2$ & $1/3$\\ 
A:b & $1$ & $4/3$ & $3/4$ & $1/2$ & $3/2$ & $1/3$\\ 
\hline \hline 
B:Y & $1$ & $2$ & $1/2$ & $1/2$ & $2$ & $1/4$\\ 
B:b & $1$ & $2$ & $1/2$ & $1/2$ & $2$ & $1/4$\\ 
\hline \hline 
C:Y & $1$ & $1$ & $1$ & $1$ & $2$ & $1/2$\\ seb
C:b & $1$ & $4/3$ & $3/4$ & $1$ & $2$ & $1/2$\\ 
\hline \hline 
D:Y & $1$ & $2$ & $1/2$ & $1$ & $2$ & $1/2$\\ 
D:b & $1$ & $2$ & $1/2$ & $1$ & $2$ & $1/2$\\ 
\hline \hline 
E:Y & $1$ & $3$ & $1/3$ & $1$ & $3$ & $1/3$\\ 
E:b & $1$ & $3$ & $1/3$ & $1$ & $3$ & $1/3$\\ 
\end{tabular} 
\end{center} 
\caption{"Dynamic" and equilibrium exponents for the Y-model (Y) and 
the  
b-model 
with bubbles (b) 
as defined in eq.(\ref{scaling}). The regimes A,B,C,D and E are those shown in Fig.1.} 
\end{table} 
} 
 
{\bf Acknowledgments:} This work was supported by MURST(COFIN 99). DM also
acknowledges INFM funding. We thank H. Dobbs and A. Trovato for discussions.
\newpage

 
\end{multicols} 

\begin{references}
\bibitem{nota1} 
Notice that along the $x$-direction the configurations are those 
of two interacting random walks which never intersect 
each other. This differs from the model considered by 
Poland and Scheraga\cite{PS66} with $f=0$, where the two walks can intersect. 
This difference seems at first subtle but it is indeed crucial: in our 
model a finite temperature transition exists 
in the other one not.
\bibitem{saws} It is interesting to notice that a diagram similar to 
the one of Fig. 2 can be obtained also if we eliminate directness 
along one direction, e.g. if we model the two DNA strands with two 
mutually and self avoiding walks\cite{unpub}.
\bibitem{note2} The critical line of this model is given by 
$f_c(T)=T \cosh^{-1}\left[\exp{\frac{1}{T}}-1 \right]$ 
and can be easily obtained  by modifying the recursion relation of     
  $d_n(x)$ in the case $x=1$.
\bibitem{not3} The exponent $\theta_1$ in regime $B$ in the b-model 
can be explained if we postulate that a waiting time $\tau$ 
 is needed for the merging  of a  
bubble into the liberated part (and also for the opposite process since at  
criticality the two processes equilibrate). The waiting time may  
reasonably be  proportional to $l(t)^2$, the time needed for the distance 
between the ends of the just opened bubble (which should obey the 
Edward-Wilkinson equation, \cite{barabasi}) to get to an ``equilibrium'' value  
$\propto l(t)^{\frac{1}{2}}$, and this gives  
$m\sim l(t)\left(\frac{t}{l(t)^2}\right)^{\frac{1}{2}}=t^{\frac{1}{2}}$.
\bibitem{note4} At $f=0$, $N\beta\frac{d{\cal F}(m)}{dm}$ displays 
  also a subleading   
 (for large $m$) entropic contribution which in $d=2$ is $-\frac{1}{2m}$. This  
should not affect the mapping proposed in the text except perhaps just 
at the critical temperature $T_m$.
\end{references}
\end{document}